\documentclass[9pt,conference]{IEEEtran}
\IEEEoverridecommandlockouts

\usepackage{booktabs}
\usepackage{tikz}
\usepackage{pgfplots}
\usepackage{multirow}
\usetikzlibrary{positioning,calc}%
\usetikzlibrary{backgrounds}
\usepackage{color, colortbl}
\usetikzlibrary{decorations.fractals}
\usetikzlibrary{patterns}
\usepackage{booktabs}
\usepackage{url}

\usepackage{cite}
\usepackage{amsmath,amssymb,amsfonts}
\usepackage{algorithmic}
\usepackage{graphicx}
\usepackage{textcomp}
\usepackage{xcolor}

\pgfplotsset{compat=1.18}

\def\BibTeX{{\rm B\kern-.05em{\sc i\kern-.025em b}\kern-.08em
    T\kern-.1667em\lower.7ex\hbox{E}\kern-.125emX}}
\begin{document}

\title{Optimizing Speech Multi-View Feature Fusion \\ through Conditional Computation
\thanks{This work was supported in part by the National Science Foundation of China (No.62276056), the Natural Science Foundation of Liaoning Province of China (2022-KF-16-01), the Fundamental Research Funds for the Central Universities (Nos. N2216016 and N2316002), the Yunnan Fundamental Research Projects (No. 202401BC070021), and the Program of Introducing Talents of Discipline to Universities, Plan 111 (No.B16009).}
\thanks{$^{\star}$Equal contribution.}
\thanks{$\dagger$Corresponding author}
}

\author{Weiqiao Shan\textsuperscript{1}, Yuhao Zhang\textsuperscript{2$^{\star}$}, Yuchen Han\textsuperscript{1}, Bei Li\textsuperscript{3}, Xiaofeng Zhao\textsuperscript{4}, Yuang Li\textsuperscript{4}, Min Zhang\textsuperscript{4},\\ Hao Yang\textsuperscript{4}, Tong Xiao\textsuperscript{1,5$\dagger$}, Jingbo Zhu\textsuperscript{1,5} \\

\IEEEauthorblockA{\textsuperscript{1}School of Computer Science and Engineering, Northeastern University, Shenyang, China\\
\textsuperscript{2}The Chinese University of Hong Kong, Shenzhen, China\\
\textsuperscript{3}Meituan, Beijing, China\\
\textsuperscript{4}Huawei Translation Services Center, Beijing, China\\
\textsuperscript{5}NiuTrans Research, Shenyang, China}}

\maketitle
\begin{abstract}

Recent advancements have highlighted the efficacy of self-supervised learning (SSL) features in various speech-related tasks, providing lightweight and versatile multi-view speech representations. However, our study reveals that while SSL features expedite model convergence, they conflict with traditional spectral features like FBanks in terms of update directions. In response, we propose a novel generalized feature fusion framework grounded in conditional computation, featuring a gradient-sensitive gating network and a multi-stage dropout strategy. This framework mitigates feature conflicts and bolsters model robustness to multi-view input features. By integrating SSL and spectral features, our approach accelerates convergence and maintains performance on par with spectral models across multiple speech translation tasks on the MUSTC dataset.

\end{abstract}


\section{Introduction}

Conventional speech processing methods rely on spectral features (e.g., MFCC, FBanks)~\cite{rabiner2007introduction, thodoroff2016learning, chan2015listen}, while recently, features learned through self-supervised learning (SSL) methods have garnered the attention of researchers~\cite{schneider2019wav2vec, baevski2022data2vec, baevski2020wav2vec}. The SSL-based features (S-feature) alleviate the variability of signals in a fixed representation space using the vector quantization (VQ) method~\cite{hsu2021hubert}. This type of feature is easy to learn and operate, and offers multi-view representations for speech processing tasks~\cite{shi2023exploration, shi2023multi}. Furthermore, the S-feature demonstrates a strong capacity for integrating contextual information, making it particularly effective for downstream tasks~\cite{hsu2021hubert,kim2023unitspeech}, even beneficial for building speech large language models~\cite{zhang2023speechgpt}.

\begin{figure}[t]
\centerline{\includegraphics[width=0.45\textwidth]{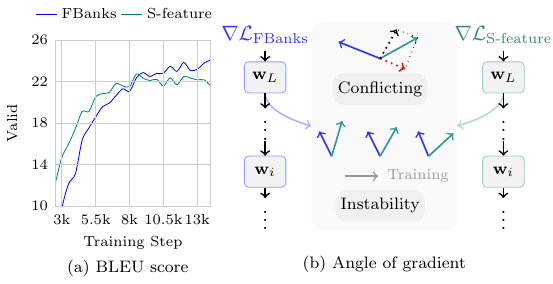}}
\caption{The BLEU scores of the model using only FBanks or S-feature (a). When we freeze all the model parameters during training and sequentially input the two features, we observe that up to 32\% gradients generated by the two features contain conflicting components, and the angle between the two gradients increases as training progresses (b).}
\label{fig:loss-acc-for-fbank-or-unit-only}
\vspace{-0.5em}
\end{figure}

Although the S-feature incorporates more in-context information learned from the pre-training stage, it suffers from information loss during the VQ process. This makes it difficult to achieve the strongest performance compared to spectral features in complicated tasks. Fig.~\ref{fig:loss-acc-for-fbank-or-unit-only}(a) shows an example in the Speech Translation (ST) task. The S-feature demonstrates rapid convergence but fails to outperform the spectral feature. Inspired by the work that fuses multi-view features to enhance the performance of downstream tasks~\cite{wan2021multi, yan2021deep, huh2024platonic}, we naturally raised the question: can we leverage the strengths of the two views of feature to improve complex speech-to-text tasks?

Though multi-view feature fusion has been achieved for low-resource cases and emotion recognition in separate studies~\cite{berrebbi2022combining, hyeon2024improving}, there remains a lack of discussion on how to accomplish this under more general conditions. Our primary analysis reveals a significant divergence in the update directions between FBanks and S-feature as shown in Fig.~\ref{fig:loss-acc-for-fbank-or-unit-only}(b). This gradient conflict explains why simple fusion methods fail to effectively integrate multi-view features (Tab.~\ref{tab:GSGN-vs-simple-gate}). To overcome this limitation, we introduce conditional computation which dynamically activates parts of the model parameters based on the inputs through a gating network~\cite{bengio2013deep}. This dynamic activation property allows the model to better adapt to the input characteristics, similar to the widely discussed early exit models~\cite{elbayad2019depth, elhoushi2024layer} and the mixture of experts used in multitask learning~\cite{ma2018modeling, gupta2022sparsely}.

Specifically, we propose a gradient-sensitive gating network (GSGN) that implements conditional computation for the gradient of parameters, enhancing the adaptability of our approach by dynamically computing the correlation between heterogeneous features and adjusting the weights of different feature branches to achieve optimal fusion. We carry on the experiences in MuST-C for three languages (En-De, En-Es, and En-Fr) in ST tasks. Our proposed method effectively resolves the conflict between the two features. For models trained from scratch, our method can provide comparable performance and average $1.24$ times training speedup compared to the model with the FBanks feature. As opposed to the pre-trained ST model, our approach is also able to provide a corresponding training acceleration effect when the S-feature is sufficiently effective and guarantees the performance of the model.

\section{Method}
\label{sec:method}

\subsection{Architecture}

We adopt an end-to-end ST system following previous work~\cite{xu2021stacked, zhang2024soft}. The system consists of an acoustic encoder (A-enc), a textual encoder (T-enc), and a decoder (dec). The forward process of the network can be denoted as follows:
\begin{align}
\mathbf{h}_{\text{A}} & = \text{A-enc}(\mathbf{x}) \nonumber \\
\mathbf{h}_{\text{T}} & = \text{T-enc}(\mathbf{h}_{\text{A}}) \nonumber \\
\mathbf{h}_{\text{out}} & = \text{dec}(\mathbf{h}_{\text{T}})
\end{align}

\begin{figure}[t]

\begin{minipage}[b]{1.0\linewidth}
  \centering
  \includegraphics[width=0.78\textwidth]{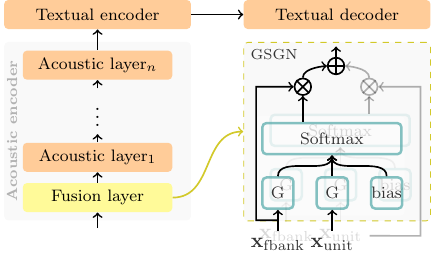}
\end{minipage}
\caption{ST model architecture.}
\label{fig:st-model-arch}
\end{figure}

During the training phase, the model generates the final prediction based on the decoder's output and obtains the final loss via the subsequent cross-entropy function:
\begin{align}
\mathcal{L} & = \text{CrossEntropy}(\mathbf{w}_{\text{out}} \mathbf{h}_{\text{out}}, \mathbf{\hat{y}})
\end{align}
where $\mathbf{\hat{y}}$ is the target text. Conventional ST tasks typically use the FBanks feature as input, denoted as $\mathbf{x}=\mathbf{x}_\text{fbank}$. In this paper, we introduce a representation with general and contextual information as a multi-view speech representation, the S-feature feature, which is represented as $\mathbf{x}_\text{unit}$, we incorporate new modules that dynamically fuse the two features through an additional fusion layer as shown in Fig.~\ref{fig:st-model-arch}. 

\subsection{Gradient-sensitive Gating Network for Conflicting Gradient}

We first present the specific formulation of our gradient-sensitive gating network. It accepts two inputs with embedding dimension size $D$, $\mathbf{x}_{\text{fbank}}, \mathbf{x}_{\text{unit}} \in \mathbb{R}^{1 \times D}$. Each input undergoes a linear mapping, resulting in the gating vector $\mathbf{g} \in \mathbb{R}^{1 \times D}$ as follows:
\begin{equation}
    \mathbf{g_{[\cdot]}} = \text{Sigmoid}(\text{Linear}_{[\cdot]}^1(\mathbf{x}_{\text{fbank}}) + \text{Linear}_{[\cdot]}^2(\mathbf{x}_{\text{unit}}) + \text{bias}_{[\cdot]})
\end{equation}
For input $\mathbf{x}_{\text{fbank}}$ and $\mathbf{x}_{\text{unit}}$, we compute the gating vector $\mathbf{g}_{\text{fbank}}$ and $\mathbf{g}_{\text{unit}}$ as the weight separately.
Finally we fuse the two features by Hadamard product:
\begin{align}
\mathbf{x}_\text{fusion} &= \mathbf{g}_{\text{fbank}} \circ \mathbf{x}_{\text{fbank}} + \mathbf{g}_{\text{unit}} \circ \mathbf{x}_{\text{unit}}
\label{eq:gate-GSGN-func}
\end{align}

Given that the network structure in use is a residual network \cite{he2016deep}, we express the transfer of each layer using the following equation:
\begin{align}
\mathbf{h}_{i + 1} = \underbrace{F(\mathbf{h}_{i})}_{\text{residual}} +\underbrace{\mathbf{h}_{i}}_{\text{shortcut}} \approx \underbrace{\mathbf{w}_{i} \mathbf{h}_{i} + \mathbf{b}_{i}}_{\text{residual}} + \underbrace{\mathbf{h}_{i}}_{\text{shortcut}}
\label{eq:layer-simple}
\end{align}
where $\mathbf{h}_i$ is the hidden state of the $i$-th layer and $F(\cdot)$ denotes the function of this layer. For simplicity, we assume a linear model for $F(\cdot)$, omitting nonlinear operations and layer normalization.

Taking the input $\mathbf{x} = \mathbf{x}_\text{fusion}$ as an example, the gradient for the parameter $\mathbf{w}_{i}$ in layer $i$ can be expressed as:

\begin{equation}
    \frac{\partial \mathcal{L}}{\partial \mathbf{w}_i} = 
\underbrace{(\mathbf{g}_{\text{fbank}} \circ \mathbf{x}_{\text{fbank}} + \mathbf{g}_{\text{unit}} \circ \mathbf{x}_{\text{unit}})^{\text{T}}}_{\text{Input-dependent}} \times \underbrace{\sum_{j=0,j\neq i}^{n}(\mathbf{w}^{\text{T}}_j + \mathbf{1}) 
 \frac{\partial \mathcal{L}}{\partial \mathbf{h}_{\text{out}}}}_{\text{Input-independent}}
\label{eq:gradient-fusion}
\end{equation}

We can derive the gradient $\frac{\partial \mathcal{L}_{\text{fbank}}}{\partial \mathbf{w}_i}$ when we only use the FBanks feature $\mathbf{x}_\text{fbank}$ to replace the input-dependent term in Eq.~\eqref{eq:gradient-fusion}. Similarly, we use $\frac{\partial \mathcal{L}_{\text{unit}}}{\partial \mathbf{w}_i}$ to denotes the gradient when $\mathbf{x}_\text{unit}$ is the only input.
Given that the Hadamard product scales the elements directly, Eq.~\eqref{eq:gradient-fusion} can be transformed into a weighted summation of two gradients when different features are input separately under frozen parameters.
\begin{align}
&\frac{\partial \mathcal{L}}{\partial \mathbf{w}_i} = \mathbf{g}_{\text{fbank}} \circ \frac{\partial \mathcal{L}_{\text{fbank}}}{\partial \mathbf{w}_i} + \mathbf{g}_{\text{unit}} \circ \frac{\partial \mathcal{L}_{\text{unit}}}{\partial \mathbf{w}_i}
\label{eq:gradient-weight-gating-weight}
\end{align}

As illustrated in Fig.~\ref{fig:loss-acc-for-fbank-or-unit-only}, the two gradients $\frac{\partial \mathcal{L}_{\text{fbank}}}{\partial w_i}$ and $\frac{\partial \mathcal{L}_{\text{unit}}}{\partial w_i}$ contain conflicting components when $\cos(\theta) < 0$. Such conflicts in gradients are commonly observed in multitask learning and can lead to the seesaw phenomenon~\cite{tang2020progressive}. To mitigate this issue, an efficient strategy is to eliminate the components of one gradient $\overrightarrow{b}$, that conflict with the gradient of the main feature $\overrightarrow{a}$~\cite{yu2020gradient}.
\begin{align}
\text{Deconflict}(\overrightarrow{a}, \overrightarrow{b}) = \overrightarrow{b} - \frac{\lVert\overrightarrow{b}\rVert \cos(\theta)}{\lVert\overrightarrow{a}\rVert}\overrightarrow{a}
\label{eq:gradient-eliminate}
\end{align}

Therefore the gradient of the model should be corrected to the summation of the two gradients without conflicting components.
\begin{align}
\begin{cases}
\overrightarrow{a} + \overrightarrow{b}, & \text{if } \cos(\theta) >= 0, \\
\overrightarrow{a} + \text{Deconflict}(\overrightarrow{a}, \overrightarrow{b}) & \text{else}
\end{cases}
\label{eq:gradient-new}
\end{align}

Based on the Eq.~\eqref{eq:gradient-eliminate}, the final gradient of the model is as follows:
\begin{align}
\begin{cases}
\textbf{1} \circ \overrightarrow{a} + \textbf{1} \circ \overrightarrow{b}, & \text{if } \cos(\theta) >= 0, \\
(\textbf{1} - \frac{\lVert\overrightarrow{b}\rVert \cos(\theta)}{\lVert\overrightarrow{a}\rVert}) \circ \overrightarrow{a} + \textbf{1} \circ \overrightarrow{b} & \text{else}
\end{cases}
\label{eq:gradient-new-final}
\end{align}

Building on the previous description, we observe that when we set $\overrightarrow{a}=\frac{\partial \mathcal{L}_{\text{fbank}}}{\partial \mathbf{w}_i}$ and $\overrightarrow{b}=\frac{\partial \mathcal{L}_{\text{unit}}}{\partial \mathbf{w}_i}$, we only need to adjust $\mathbf{g}_{[\cdot]}$ in Eq.~\eqref{eq:gradient-weight-gating-weight} to satisfy Eq.~\eqref{eq:gradient-new-final}, and in turn eliminate the conflicting components of the gradients produced by the two features. Consequently, we can derive the ideal gradient by constraining $\mathbf{g}_{\text{unit}}=\textbf{1}$ and introducing an additional loss term for $\mathbf{g}_{\text{fbank}}$.
\begin{align}
& \mathcal{L}_{\text{gate}}  = \nonumber \\
&\begin{cases}
\text{MSE}(\mathbf{g}_{\text{fbank}}, \textbf{1}) & \text{if } \cos(\theta) >= 0, \\
\text{MSE}(\mathbf{g}_{\text{fbank}}, (\textbf{1} - \frac{\lVert\frac{\partial \mathcal{L}_{\text{unit}}}{\partial \mathbf{w}_i}\rVert \cos(\theta)}{\lVert\frac{\partial \mathcal{L}_{\text{fbank}}}{\partial \mathbf{w}_i}\rVert})) & \text{else}
\end{cases}
\end{align}

So the final loss of the model is:
\begin{align}
\mathcal{L}_{\text{final}} = \mathcal{L} + \mathcal{L}_{\text{gate}}
\end{align}

\begin{table*}[t]
\small
\caption{Main result of our method.}
\centering
\begin{tabular}{l|ll|ll|ll}
\toprule
\multirow{2}{*}{Model} & \multicolumn{2}{c|}{En-De} & \multicolumn{2}{c|}{En-Fr} & \multicolumn{2}{c}{En-Es} \\
 & Test & Epoch & Test & Epoch & Test & Epoch \\ 
\midrule
S2T(FBanks-only) & 25.08 & 40 & 36.22 & 35 & 30.04 & 37 \\
S2T(S-feature-only) & 23.33 & 27 & 33.00 & 25 & 26.52 & 23 \\
\midrule
\rowcolor{black!10} S2T-GSGN & 24.18 & 28 & 33.91 & 25 & 28.61 & 24 \\
\rowcolor{black!10} S2T-GSGN+Drop & 25.26 (\textcolor{green!70!black}{+0.18}) & 30 (\textcolor{green!70!black}{$\times1.33$}) & 36.24 (\textcolor{green!70!black}{+0.02}) & 30 (\textcolor{green!70!black}{$\times1.17$}) & 30.24 (\textcolor{green!70!black}{+0.20}) & 30 (\textcolor{green!70!black}{$\times1.23$}) \\
\bottomrule
\end{tabular}
\label{tab:main-exp-for-GSGN-drop}
\end{table*}


\subsection{Multi-stage dropout for Instability Gradient}
For the instability gradient shown in Fig.~\ref{fig:loss-acc-for-fbank-or-unit-only}, the $\cos(\theta)$ between the two gradients from each feature is not constant during training. The variation in gradient directions highlights the inconsistency in the roles of different features during the training process. Initially, the roles of these features may be highly similar. However, these updating directions change significantly during the training process, suggesting that the branches adapt to focus on different aspects of the data at various stages.

To enhance the model's robustness to each input feature and ensure that all features can be fully utilized, we dynamically sample from three features during the training stage and input them into the next layer. Specifically, we sample the FBanks, S-feature, and fusion features with two fixed threshold $\delta_{\text{fbank}}, \delta_{\text{unit}}$ and a random probability $p \in (0,1)$.
\begin{align}
\mathbf{x}=
\begin{cases}
\mathbf{x}_{\text{fbank}}, & \text{when} \quad p < \delta_{\text{fbank}} \\
\mathbf{x}_{\text{unit}}, & \text{when} \quad \delta_{\text{fbank}} <= p < \delta_{\text{fbank}} + \delta_{\text{unit}} \\
\mathbf{x}_{\text{fusion}}, & \text{when} \quad \delta_{\text{fbank}} + \delta_{\text{unit}} <= p
\end{cases}
\end{align}

\section{Experiments}
\label{sec:experiments}

\subsection{Datasets}

We mainly experimented on the MuST-C dataset (MuSTC) including three benchmarks~\cite{di2019must, zhang2024soft}, English-German
(En-De), English-French (En-Fr) and English-Spanish (En-Es), and we report results on the test sets for all models. For the experiment based on the pre-trained model, we obtain the pre-trained ASR model on the LibriSpeech~\cite{panayotov2015librispeech} dataset to initialize the acoustic encoder. For pre-training textual encoder and decoder, pre-training data include WMT16, WMT14, and WMT13 for En-De, En-Fr, and En-Es three tasks respectively. The detailed training setup throughout the pre-training phase follows~\cite{han2023modality}. To obtain the S-feature, we use the pre-trained and fine-tuned ASR Hubert models~\cite{hsu2021hubert}, and setting 500 for $k$-means cluster center based on the release model\footnote{https://dl.fbaipublicfiles.com/hubert/hubert\_base\_ls960\_L9\_km500.bin}.

\subsection{Experimental Settings}

Based on the Transformer model~\cite{vaswani2017attention}, our model consists of an acoustic encoder with 12 layers, and a textual encoder and decoder with 6 layers. We set the model hidden size to 512 and the feedforward size to 2048, with 8 head attention. We applied dropout with a value of 0.1 to attention weights and 0.2 to residual connections, respectively. We also used label smoothing of 0.1 to handle overfitting. The Adam ($\beta_1$ = 0.9, $\beta_2$ = 0.98) optimizer with a warmup step of 4K was adopted. Additionally, we use the early stopping strategy set the patience = 10.

For GSGN, we follow the setting in Eq.~\ref{eq:gate-GSGN-func} to compute a weight matrix $\mathbf{g_{[\cdot]}}$. The matrix $\mathbf{g_{[\cdot]}}$ has the same dimension with each input feature, where each $\mathbf{g_{[\cdot]}}$ is given by $\text{Linear}_{[\cdot]}: \mathbb{R}^{1 \times D} \to \mathbb{R}^{1 \times D}$, and we set D to 512 following the model hidden size.

For multi-stage dropout, we empirically divide the training process into three stages, when the current $\text{epoch} < 10$, we set the $\delta_{\text{fbank}} = 0.3, \delta_{\text{unit}} = 0$ for better performance of gradient descent. During the next stage, when $10 \leq \text{epoch} < 25$ we increase the $\delta_{\text{fbank}} = 0.5, \delta_{\text{unit}} = 0.3$ to ensure the model fully leverages the different features. Finally, for $25 \leq \text{epoch}$ and we revert the thresholds to $\delta_{\text{fbank}} = 0.3, \delta_{\text{unit}} = 0$, allowing the fusion branch to achieve better performance during the inference stage.

During the inference stage, we average the best 10 checkpoints on the valid set for evaluation with beam size = 5 for beam search, and length penalty = 1.0, we provide the last epoch at the average 10 checkpoints to indicate the convergence speed. We do a down-sample for the S-feature for the same length as the FBanks feature, and we use the same setting as the third stage in multi-stage dropout to sample from three branches. We evaluate translation quality with sacreBLEU~\cite{post2018call}.

\subsection{Results}

Tab.~\ref{tab:main-exp-for-GSGN-drop} shows the performance of models based on GSGN and GSGN+Drop. The GSGN+Drop method achieves better results and faster convergence compared to the baseline model. We find that the GSGN method effectively improves performance while maintaining a comparable convergence rate to the model with the S-feature. Furthermore, GSGN+Drop enhances the robustness of the model to the two input features while maintaining an acceptable training speedup\footnote{Our code is available at https://github.com/shanweiqiao/GSGN.git}.

\section{ANALYSIS}

\subsection{Effectiveness of S-feature}

\begin{table}[t]
\small
\caption{Random noise experiment.}
\centering
\begin{tabular}{l|ccc}
\toprule
Noise & Model & Test & Epoch \\
\midrule
Sum & S2T(FBanks-only) & 25.37 & 38 \\
Replace & S2T-GSGN & 25.19 & 37 \\
Replace & S2T-GSGN+Drop & 25.3 & 39 \\
\bottomrule
\end{tabular}
\label{tab:rand-noise-for-unit}
\end{table}

To demonstrate the effectiveness of the S-feature, we incorporated random noise into the model in the EN-DE direction. The random noise sampling from a uniform distribution has the same maximum and minimum values as the S-feature across the entire training set, the results are shown in Tab.~\ref{tab:rand-noise-for-unit}. We employed two noise addition methods 1)Sum: adding noise directly to the current input, and 2)Replace: replacing the S-feature with noise. While the model with random noise achieves better results, it requires more training epochs. This suggests that even adding noise directly to the baseline model with FBanks can enhance performance by improving robustness and mitigating overfitting. When replacing the S-feature with random noise in the S2T-GSGN model, we observed better performance but slower convergence, indicating that the S-feature indeed provides a fast convergence capability for the ST model, rather than behaving like inaccurate noise. Furthermore, in the S2T-GSGN+Drop model, the results suggest that the multi-stage dropout method enhances the model's ability to leverage different features, rather than solely serving as a robust training method.

\subsection{Effectiveness of GSGN}

\begin{table}[t]
\small
\caption{Concatenation-based gating network vs GSGN.}
\centering
\begin{tabular}{l|cc}
\toprule
En-De & Test & Epoch \\
\midrule
S2T-Concat & 23.8 & 32\\
S2T-GSGN & 24.18 & 28\\
\bottomrule
\end{tabular}
\label{tab:GSGN-vs-simple-gate}
\end{table}

\begin{table}[t]
\small
\centering
\caption{Experiments of our method on the pre-trained models.}
\begin{tabular}{l|ll}
\toprule
En-De & Test & Epoch \\
\midrule
S2T(FBanks-only) & 27.98 & 33 \\
S2T(S-feature-only) & 25.60 & 25 \\
\midrule
S2T-GSGN & 27.61 & 25 \\
S2T-GSGN+Drop & 28.19 (\textcolor{green!70!black}{+0.21}) & 27 (\textcolor{green!70!black}{$\times1.22$}) \\
\bottomrule
\end{tabular}
\label{tab:main-exp-for-GSGN-drop-pretrain}
\end{table}

We also investigated the effectiveness of GSGN by comparing a simple concatenation-based gating network on En-De direction based on the model trained from scratch. The new gating concat two features $\mathbf{x}_\text{concat}=[\mathbf{x}_\text{fbank};\mathbf{x}_\text{unit}] \in \mathbb{R}^{1 \times 2D}$ and maps them back to original dimensions by  $\text{Linear}(\mathbf{x}_\text{concat}): \mathbb{R}^{1 \times 2D} \to \mathbb{R}^{1 \times D}$. We find that GSGN makes more efficient use of the S-feature and achieves faster and better convergence as shown in Tab.~\ref{tab:GSGN-vs-simple-gate}. This suggests that the simple concatenation-based gating network fails to resolve the conflicting gradient problem between two features, thereby limiting its ability to fuse them effectively.

\subsection{Effectiveness under the pre-trained model}

\begin{figure}[t]
\centerline{\includegraphics[width=0.40\textwidth]{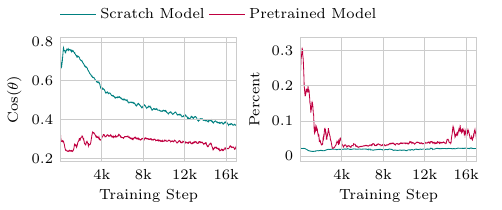}}
\caption{The $\cos(\theta)$ between two gradients generated by each feature. We find \textbf{Instability}(left): The $\cos(\theta)$ becomes smaller and smaller with training for $\cos(\theta) > 0$, and \textbf{Conflicting}(right): Percent of conflicting gradients($\cos(\theta) < 0$) among all gradients in an input batch.}
\label{fig:grad-conflict-and-instability-data}
\end{figure}

We conducted the same experiment on the ST model initialized with a pre-trained model, as shown in Tab.~\ref{tab:main-exp-for-GSGN-drop-pretrain}. The results indicate that the GSGN is more effective on the pre-trained models than on the models trained from scratch. We find that this difference is due to the varying frequencies of gradient conflicts and instability in the trained from scratch and pre-trained models (Fig.~\ref{fig:grad-conflict-and-instability-data}). In pre-trained models, gradient conflict is particularly common at the beginning of training, and GSGN effectively mitigates this issue, leading to improved results. In contrast, gradient instability is more prominent in models trained from scratch, making the multi-stage dropout method more beneficial.

\subsection{Weight Analysis for GSGN}

\begin{figure}[t]
  \centering
  \centerline{\includegraphics[width=0.40\textwidth]{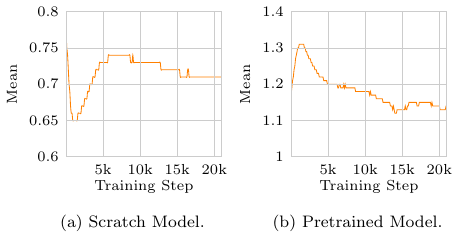}}  
  \caption{The mean value of $\mathbf{g}_{\text{fbank}}$ in the model training from scratch (left) and the pre-trained model (right).}
\label{fig:fbank-gate-weight}
\end{figure}

Additionally, we analyzed the weights assigned by the GSGN to the FBanks features in the S2T-GSGN+Drop model. As shown in Fig.~\ref{fig:fbank-gate-weight}, for the model trained from scratch, we observed that the GSGN pays more attention to the S-feature in the initial phase of the training to leverage the rapid convergence properties. During training, GSGN gradually fuses more FBanks to achieve a more stable and effective descent process. 

For the pre-trained model, we observe that GSGN assigns a higher weight to $\mathbf{g}_{\text{fbank}}$. This is because, when gradient conflicts occur, the $\cos(\theta)$ is less than 0, and the term $(\textbf{1} - \frac{\lVert\overrightarrow{b}\rVert \cos(\theta)}{\lVert\overrightarrow{a}\rVert})$ in Eq.~\ref{eq:gradient-new-final}, or equivalently the mean of $\mathbf{g}_{\text{fbank}}$ in Eq.~\ref{eq:gate-GSGN-func} would be greater than 1. We find that 89.53\% of the elements in $\mathbf{g}_{\text{fbank}}$ exceed 1 in the pre-trained model. This indicates that the conflict gradient in the pre-trained model is severe, which aligns with our observation in Fig.~\ref{fig:grad-conflict-and-instability-data} (right).

\begin{figure}[htp]
  \centering
  \centerline{\includegraphics[width=0.20\textwidth]{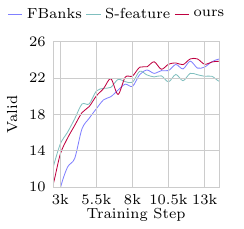}}  
  \caption{The final result of our method.}
\label{fig:final-result}
\end{figure}

It is important to emphasize that both instability gradient and conflict gradient co-exist in models trained from scratch and pre-trained models. Although the conflict gradient is particularly common in the pre-trained models, there are still 10.46\% of the elements in $\mathbf{g}_{\text{fbank}}$ that are less than 1. Conversely, in models trained from scratch, about 8.63\% of the elements in $\mathbf{g}_{\text{fbank}}$ exceed 1. This demonstrates that GSGN, in combination with the multi-stage dropout method, effectively addresses the gradient-related issues in both cases. After correctly handling the gradient problem, we find that our method achieves faster convergence speedup compared to the model using only FBanks, and superior performance compared to the model using only the S-feature. Our fusion method successfully leverages the strengths of both features, as shown in Fig.~\ref{fig:final-result}.

\section{Conclusion and Future Work}
\label{sec:conclusion}

Recent studies have demonstrated the effectiveness of S-features across various speech-related tasks, offering lightweight and general-purpose multi-view speech representations. However, we observed that while S-features exhibit rapid convergence, they fail to outperform FBanks features due to conflicts in their update directions. To address this, we propose a conditional computation method with a specialized gradient-sensitive gating network, which dynamically computes the correlation between heterogeneous features and adjusts the weights of different feature branches. Our method effectively mitigates conflicts between the two features while exploiting their complementary aspects. As a result, we achieve comparable performance and average $1.24$ times training speedup compared to models using only FBanks features, for both models trained from scratch and pre-trained models in the MuST-C En-De, En-Es, and En-Fr speech translation tasks. In the future, we aim to explore more effective fusion approaches by incorporating varied unit representations from more pre-trained models, as well as exploiting the pre-trained models directly.

\vfill\pagebreak

\newpage

\bibliographystyle{IEEEbib}
\bibliography{strings}

\end{document}